# Terahertz antiferromagnetic dynamics induced by ultrafast spin currents

Sanjay René[1], Artem Levchuk[1], Amr Abdelsamie[2,3], Zixin Li[1], Pauline Dufour[2], Arthur Chaudron[2], Florian Godel[2], Jean-Baptiste Moussy[1], Karim Bouzehouane[2], Stéphane Fusil[2,4], Vincent Garcia[2], Michel Viret[1], Jean-Yves Chauleau[1,*]

[1]SPEC, CEA-Saclay, CNRS, UMR3680 , Université Paris-Saclay, 91191 Gif-sur-Yvette, France
[2]Laboratoire Albert Fert, CNRS, Thales, Université Paris-Saclay, 91767 Palaiseau, France
[3]Laboratoire Charles Coulomb, Université de Montpellier and CNRS, 34095 Montpellier, France
[4]Université d'Evry, Université Paris-Saclay, 91000 Evry, France

*jean-yves.chauleau@cea.fr

**Insulating antiferromagnets are anticipated as the main protagonists of ultrafast spintronics, with their intrinsic terahertz dynamics and their abililty to transport spin information over long distances. However, direct transfer of spin angular momentum to an antiferromagnetic insulator at picosecond time scales remains to be demonstrated. Here, studying the ultrafast behaviour of ferromagnetic metal/antiferromagnetic insulator bilayers, we evidence the generation of coherent excitations in the antiferromagnet combined with a modulation of the demagnetization behavior of the ferromagnet. This confirms that magnetic information can indeed be propagated into antiferromagnetic spin waves at picosecond timescales, thereby opening an avenue towards ultrafast manipulation of magnetic information.**

Antiferromagnets[1] (AF), in which neighbouring atomic magnetic moments are antiparallelly aligned, are now being considered as active components of future applications beyond exchange bias and magnetic pinning in spin valves. Indeed, for the past decade, antiferromagnets have become key players in spintronics[2,3], driven by efficient interactions between AF spin textures and spin currents, i.e., currents carrying spin-angular momentum. A particular attention has been devoted to insulating AF materials as they are expected to transport spin information at low-energy cost and very high frequencies. Indeed, AFs exhibit superior dynamics as their magnetic resonances[4,5] are intrinsically in the terahertz[6,7] (THz) range and the intrinsic damping of insulating compounds can be substantially low. This implies that their magnetic textures can be naturally actuated at picosecond timescales. Mastering and extending the spintronic concepts to sub-picosecond timescales using AF terahertz dynamics would pave the way for tomorrow's energy-efficient ultrafast devices.

A first AF spintronic fundamental concept is the transport of spin information through AF insulators[8–12]. This has been demonstrated in both thin films and single crystals from DC to GHz, which is far from their natural sub-THz and THz resonance frequencies. These pioneering studies have shown that spin-fluctuations in the AF can potentially propagate spin-angular momentum as evanescent waves in AF insulators[13]. The propagation may occur over long distances depending on the characteristics of the linear AF modes[14] and the size and quality of the samples involved. A second important spintronic notion, the so-called spin pumping[15] where spin currents can be dynamically generated by magnetic resonance, has been theoretically discussed in typical heavy metal/insulating AF bilayers[16]. Pioneering experiments[17,18] reported AF spin-pumping at sub-THz frequencies, emphasizing the ultrafast performances of AF spin textures. In the same vein, Rongione et al.[19] have shown that this can be extended to higher frequencies by measuring the THz emission of Pt/NiO bilayers, NiO being a prototypical insulating antiferromagnet.

Nevertheless, direct evidence of spin angular momentum transfer or spin transfer torque (STT) to an insulating antiferromagnet at picosecond and sub-picosecond timescales is still missing. Indeed, because of vanishing net magnetizations in AF textures, direct measurements of AF dynamics or AF magnons is extremely challenging. Here we tackle this issue by measuring the ultrafast dynamics of ferromagnetic/insulating AF bilayers excited by an intense femtosecond optical pulse and probed using ultrafast time-resolved magneto-optics. The strategy is to assess the effect of an exchange of spin angular momentum with the insulating AF texture on the ultrafast demagnetization characteristics of the ferromagnetic layer and detect the triggered THz dynamics in the AF layer (Figure 1a).

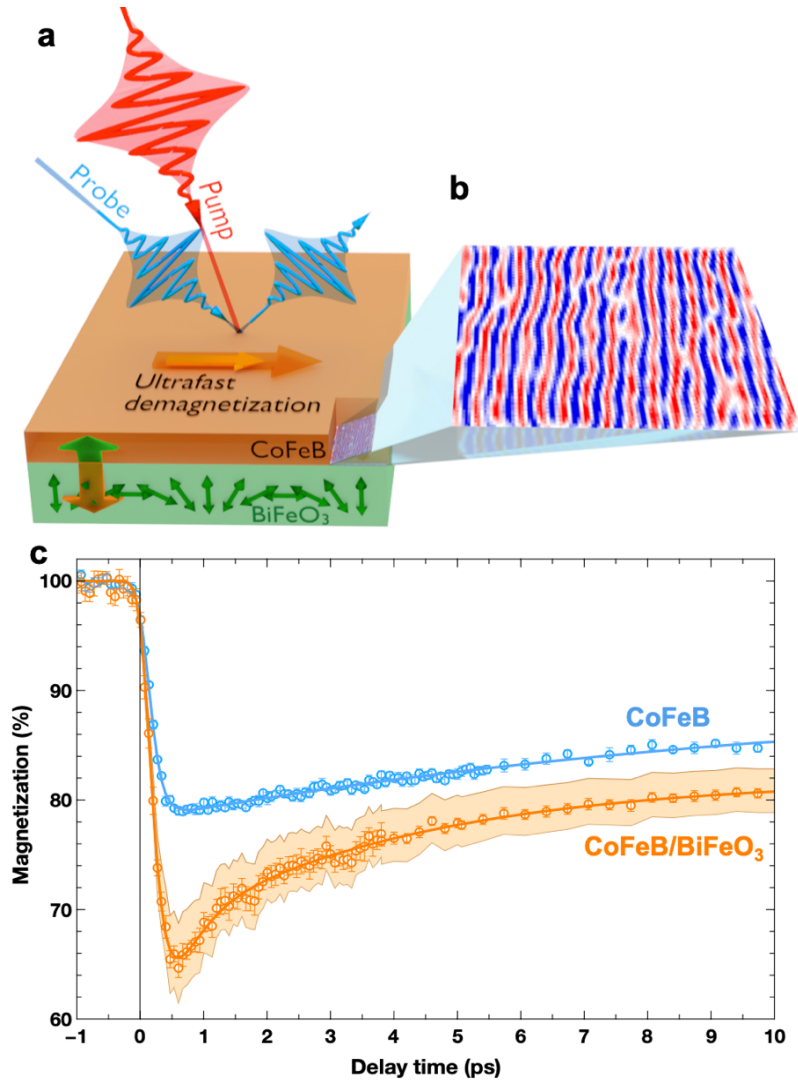

**Fig. 1**: **a**, Schematic representation of the Tr-MOKE experimental configuration. The large bi-color double arrow stands for the ultrafast exchanges of angular momentum between the ferromagnet and the antiferromagnet. **b**, Scanning nitrogen-vacancy magnetometry image of the single AF cycloidal state present in the epitaxial $BiFeO_3$ layer with a typical period of 67 nm. **c**, Ultrafast magnetization dynamics of the $CoFeB/BiFeO_3$ bilayer (orange dots) and the CoFeB reference layer (blue dots). The CoFeB magnetization and the probe incidence plane are along the $BiFeO_3$ cycloid plane or the $DyScO_3$ a-axis and the probe is p-polarized. The magnetization dynamics is presented as $\Delta M/M_0$ in percent (%) of variation.

Indeed, since its experimental evidence in 1996 by Beaurepaire et al.[20], the underlying mechanisms of laser-induced magnetization quench on sub-picosecond timescales have been extensively studied[21–24]. Although open questions still remain, various phenomena at play have been identified. First of all, scattering processes involving spin-flip events[25] transfer angular momentum to other reservoirs such as phonons. The main other class of processes involved in the ultrafast magnetization dynamics is spin transport[26–29] as ultrafast spin currents can circulate between the ferromagnetic and adjacent layers. These spin currents can be used, for example, to induce spin dynamics into a second ferromagnetic layer[30], or to produce a transient charge current in a large spin-orbit coupling material, as spins are converted into charge, triggering a terahertz radiation[31–33]. The ferromagnetic layer can be therefore considered as an ultrafast spin current generator. Interestingly, any kind of exchange of spin angular momentum with an adjacent insulating AF layer is expected to modify the intrinsic ultrafast demagnetization characteristics of the ferromagnet. In this article, we measure the influence of an adjacent insulating antiferromagnet: $BiFeO_3$ thin film on the demagnetization dynamics of an ultrathin metallic CoFeB and evidence the THz dynamics in the antiferromagnet triggered by ultrafast STT.

For the purpose of this work, an AF material with a well-established resonance in the THz range and a Néel temperature ($T_N$) far above room temperature is required. Indeed, any significant sensitivity to thermal effects would hinder the observation of an additional transfer of angular momentum. While prototypical NiO is a good candidate, with a 1 THz AF resonance and $T_N$ = 523 K (in bulk crystals), controlling its AF microscopic domain configurations in thin epitaxial layers remains challenging. Instead, the magneto-electric antiferromagnet $BiFeO_3$ (Ref. [34]) with AF electromagnons between 0.5 and 1 THz[35] and $T_N$ = 650 K is the material of choice for our study. The most salient consequence of this magneto-electric coupling is the stabilization of an AF spin texture in the form of a well characterized spin cycloid[34]. In thin films, the AF cycloidal state can be controlled by epitaxial strain engineering[36,37] and recently, $BiFeO_3$ layers harboring a single AF cycloidal domain (along with a single ferroelectric domain) have been produced[38] (Fig. 1b, Methods). Because the antiferromagnetic spin texture is homogeneous throughout the whole sample, the complexity of averaging multiple AF domains is avoided and angular variations can also bring important extra information. Besides, the polar nature of the material brings another significant advantage for the detection of the ultrafast dynamics of the AF spin texture. In $BiFeO_3$, thanks to the magneto-electric interaction, the AF dynamics triggers electro-magnons which contain a polar dynamical contribution, allowing us to experimentally demonstrate the coherent ultrafast AF response triggered by spin transfer torque.

The system's dynamics at the picosecond and sub-picosecond timescales is assessed using a time-resolved magneto-optical Kerr effect (Tr-MOKE) pump-probe setup. Fig. 1c shows the ultrafast change in magnetization as a function of time on both the reference CoFeB and CoFeB/$BiFeO_3$ layers. First of all, while our time resolution does not allow for the observation of a substantial difference in the demagnetization rates of the two systems (≈ 200 fs), a significant enhancement of the demagnetization amplitude is observed in the CoFeB/BiFeO$_3$ bilayer. Indeed, for an equivalent absorbed pump energy density in the CoFeB layer, the demagnetization amplitude of about 20% in the single CoFeB reference layer is substantially increased to 35% in the CoFeB/BiFeO$_3$ bilayer (Fig. 1c). This is a clear signature of the opening of an additional ultrafast dissipation channel due to the adjacent BiFeO$_3$ layer. A precise estimation of the demagnetization amplitudes is challenging since, despite thorough measurements of the pump fluences, the determination of the exact power absorbed in the CoFeB layer remains a delicate task (details in Supplementary Note 2). However, our estimation is that for a given incident pump, about half of the power is absorbed in the ferromagnetic layer in the case of CoFeB/BiFeO$_3$ with respect to the reference CoFeB. The second striking difference appears in the remagnetization processes. For the reference CoFeB layer, the remagnetization behavior is reminiscent of a unique spin-to-phonon dissipation channel with a characteristic time of 10 ps. In contrast, the remagnetization of the CoFeB/BiFeO$_3$ bilayer can only be understood by considering two remagnetization processes, with characteristic times of about 0.5 ps and 6 ps. The latter has the same phononic origin as that of the single CoFeB layer albeit with a slightly shorter characteristic time denoting a stronger coupling to the lattice due to the addition of BiFeO$_3$. Importantly, the short timescale remagnetization contribution evidences the opening of an additional angular momentum dissipation channel strongly that enhances the damping of the high-energy magnons generated during the demagnetization process. This is the signature of the ultrafast pumping of a spin current to the AF insulator. Although this set of measurements unambiguously demonstrates that the contact with BiFeO$_3$ opens a new dissipation channel, its possible phononic nature cannot be completely ruled out[39]. Nevertheless, magnetic coupling between two magnetic systems should be stronger than phononic mechanisms. Furthermore, angular momentum exchange should be enhanced when the frequency of the generated magnons matches that of the available AF spin wave modes, i.e., in the THz range, although incoherent spin fluctuations in the AF layer can also play a role.

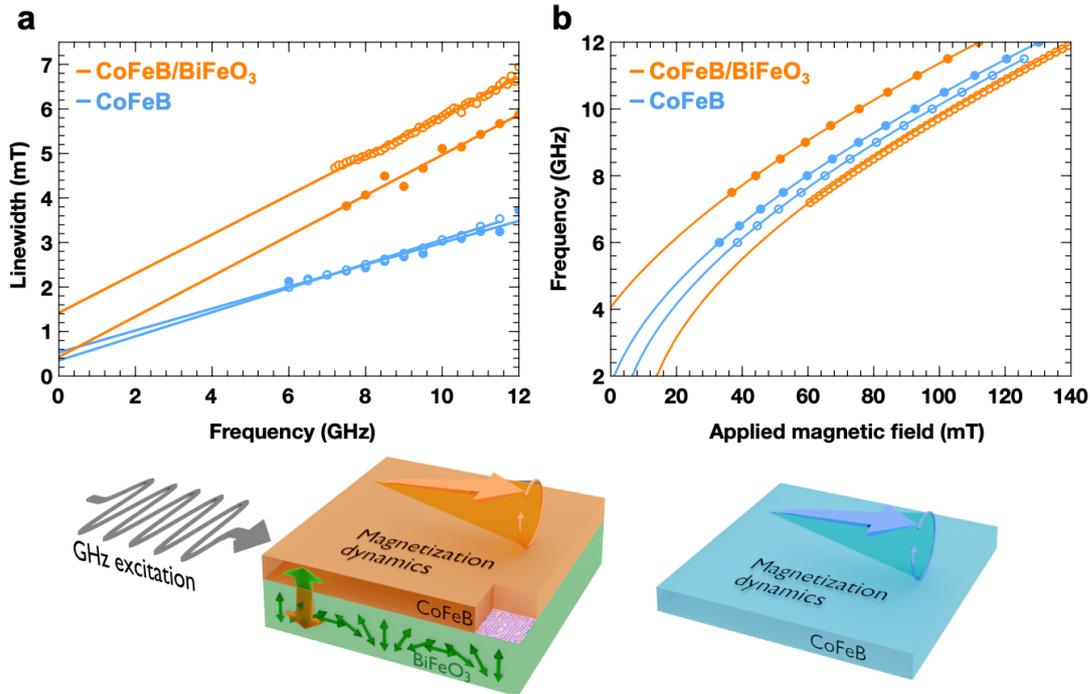

**Fig. 2** : Ferromagnetic resonance measurements performed on the CoFeB reference (blue dots) and the CoFeB/BiFeO$_3$ sample (orange dots). **a**, Frequency dependence of the linewidth ($\mu_0 \Delta H$). **b**, Field dependence of the resonance frequency. For each sample, two configurations are considered with the external magnetic field applied either along (opened dots) or perpendicular (full dots) to the a-axis of the DyScO$_3$ substrate (also, parallel to the propagation direction the BiFeO$_3$ lcycloid). The sketches represent the experimental configurations, with the large double bi-color arrow representingthe GHz and DC angular momentum exchanges between the ferromagnet and the antiferromagnet.

This incoherent channel can also be probed far from the THz AF resonances using ferromagnetic resonance (FMR), which can measure the additional magnetic dissipation induced in the ferromagnet because of the proximity with the antiferromagnet. Indeed, the magnetic damping provides a clear signature of the propensity of the AF order to absorb the angular momentum generated during ferromagnetic resonance[40]. Figure 2 shows measurements performed on both the CoFeB/BiFeO$_3$ bilayer and the CoFeB reference layer. Fig. 2a compares the FMR linewidth as a function of the excitation frequency ($f$) for the two samples. They both exhibit characteristic linear behaviours from which one can infer the magnetic damping coefficient ($\alpha$) and the inhomogeneous contribution ($\mu_0 \Delta H_0$), following: $\mu_0 \Delta H = 2\pi (\alpha/\gamma) \cdot f + \mu_0 \Delta H_0$ where $\gamma$ is the gyromagnetic ratio and $\mu_0$ the vacuum permeability. While the magnetic damping coefficient for the CoFeB reference layer is $\alpha_{\text{ref}} = 7.5 \times 10^{-3}$, the one of the CoFeB/BiFeO$_3$ bilayer is $\alpha_{\text{BFO}} = 13 \times 10^{-3}$, implying that $\alpha_{\text{BFO}} = \alpha_{\text{ref}} + \Delta\alpha$. This additional $\Delta\alpha = 5.5 \times 10^{-3}$ contribution to the damping can also be seen as a spin pumping effect where spin-angular momentum is dissipated via an additional channel mediated by antiferromagnetic fluctuations. This is consistent with what has been observed[8,40] in other ferromagnet/AF bilayers. Here, because our BiFeO$_3$ is single domain, angular measurements can provide more information on the angular momentum absorption process. While the CoFeB reference layer shows a purely isotropic behaviour, the damping and the resonance field of the CoFeB/BiFeO$_3$ system clearly depend on the direction of the applied magnetic field relative to the spin cycloid propagation direction. This is also clearly visible in Fig. 2b which shows the frequency-field dependence of the resonance for both samples, where a frequency difference is observed depending on the direction of the applied magnetic field, although with a much smaller magnitude for the reference sample. The experimental observations can be fitted by a standard Kittel formula (Methods, Supplementary Note 1), including a magneto-crystalline uniaxial anisotropy contribution ($H_K = -2K/\mu_0 M_S$) where $M_S$ is the saturation magnetization. For the reference layer, a small uniaxial anisotropy ($\mu_0 H_K \sim 2 - 3$ mT) is observed whereas the contact with the AF layer in CoFeB/BiFeO$_3$ significantly increases the anisotropy ($\mu_0 H_K \sim 11 - 16$ mT). In both cases, the easy axis is perpendicular to the DyScO$_3$ a-axis, i.e., perpendicular to the AF cycloidal plane in the case of the CoFeB/BiFeO$_3$ bilayer. This enhancement of the magneto-crystalline anisotropy is a signature of the exchange coupling[41,42] often reported at the ferromagnet/AF interface. We also find that the intrinsic magnetic damping is isotropic but the inhomogeneous contribution to the linewidth ($\mu_0\Delta H_0$) is significantly larger when the applied field is along the cycloidal direction (Fig. 2a), while the intrinsic damping is isotropic. Note that when the magnetization M is perpendicular to the cycloid, the inhomogeneous damping is close to that of the reference layer, attesting a similar quality for the CoFeB in both systems. Besides, the isotropic character of the damping points to an AF spin fluctuation origin for the additional dissipation mechanism, rather than coherent excitations of AF dynamics even far from their resonance frequencies. Indeed, as the BiFeO$_3$ layer hosts one cycloidal domain, noticeable angular variations could be expected in the coherent picture: when the injected angular momentum is perpendicular to the AF cycloidal plane, the STT is equivalent on all the magnetic moments, while, when the angular momentum is parallel to the cycloidal plane, the STT follows a sinusoidal dependence along the magnetic moments of the cycloid, leading to a lower $\Delta\alpha$.

Consequently, the isotropy of the damping indicates that at the nanosecond timescale, incoherent spin fluctuations in the AF are the main cause for the angular momentum exchange. The observations in Fig. 1c result from similar mechanisms but on a different timescale.

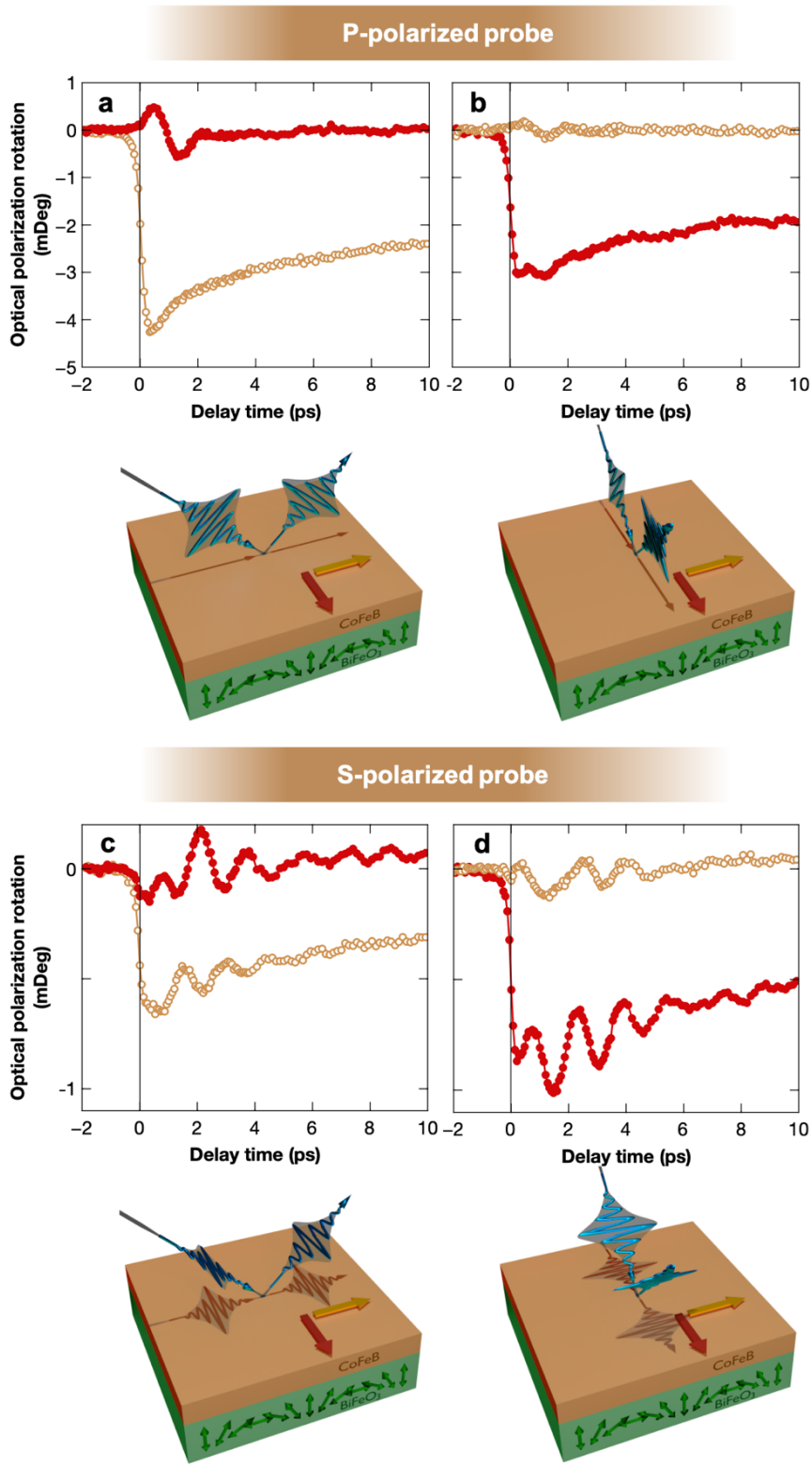

**Fig. 3** : Comparaison of the CoFeB/BiFeO$_3$ bilayer ultrafast dynamics with respect to the probe polarizations: **a**- **b**, p-polarized probe **c-d**, s-polarized probe. The plane of incidence of the probe beam and that of the BiFeO$_3$ cycloidal are, either coinciding (**a** & **c**) or are being orthogonal (**b** & **d**). For each panel, two colour codes for the data define two experimental

configurations with the CoFeB magnetization either parallel (orange) or perpendicular (dark red) to the propagation vector of the $BiFeO_3$ cycloid, as defined in the corresponding sketches.

At the picosecond however, the situation can be quite different as this timescale matches the resonant excitations of antiferromagnetic magnons. To evidence a possible coherent triggering, one needs to access the AF dynamics within the $BiFeO_3$ layer. On general grounds, this is an extremely challenging measurement, if not impossible. In the present study, we chose $BiFeO_3$ as the AF material because it can be synthesized in a single AF cycloidal state with the important extra property of being a magnetoelectric material. Indeed, this confers a polar character to the AF dynamics (see Supplementary Note 4), which offers the possibility to probe this signal through the birefringence. We thus measure the ultrafast dynamics of the $CoFeB/BiFeO_3$ bilayer with the in-plane magnetization of the CoFeB layer set either perpendicularly or parallel to the AF cycloidal plane. This can be achieved either by rotating the sample by 90° in a fixed DC magnetic field or by rotating the field to set the CoFeB magnetization transversally to the incident probe pulse. The former geometry is that of the traditional longitudinal MOKE magnetization measurement. The latter configuration only measures the transverse MOKE leading to a much smaller effect on the reflected light amplitude where, the ultrafast demagnetization in the CoFeB layer is not visible and the main contribution, at short timescales, originates from the $BiFeO_3$ birefringence. This is also highly dependent on the relative direction between the probe/pump light propagations and the crystal lattice, so one cannot rely on rotating the sample to isolate the STT-induced dynamics per se. On the other hand, unlike the CoFeB magnetization, birefringence is expected to sensitively depend on the polarization state of the probe (either p-polarized or s-polarized), suggesting that any pronounced dependence on the probe polarization would be the signature of $BiFeO_3$ dynamics.

The four possible magneto-optical configurations are gathered in Figure 3. All measurments clearly evidence some THz dynamics. The s-polarized probe optimally shows the emergence of a signal at about 0.6 THz (Fig. 3c-d), in agreement with typical electromagnon frequencies in $BiFeO_3$[35] and clearly depending on the relative orientation between the injected spin-current and the AF cycloidal plane. We recall here that in order to exclude any possible parasitic optical effects (Supplementary Figure 3), the extracted signals are always the difference between opposite magnetization directions, which is a standard approach in Tr-MOKE studies. In the present case, this means that the phase of the observed sub-THz oscillations is odd with respect to the direction of the CoFeB magnetization, i.e., it follows the direction of the induced STT.

These observations demonstrate the paramount importance of the relative angle between the injected spin-current and the antiferromagnetic cycloidal plane, emphasizing the ultrafast STT origin of the triggered AF dynamics. Direct optical pumping in the $BiFeO_3$ layer is discarded by the observation that the pump polarization has no influence on the observed dynamics (Supplementary Figure 4) and considering as well that most of the pump power is absorbed in the CoFeB layer. The experimental assessment of the exact nature of the STT-induced AF dynamics from the measured dynamical birefringence is challenging.

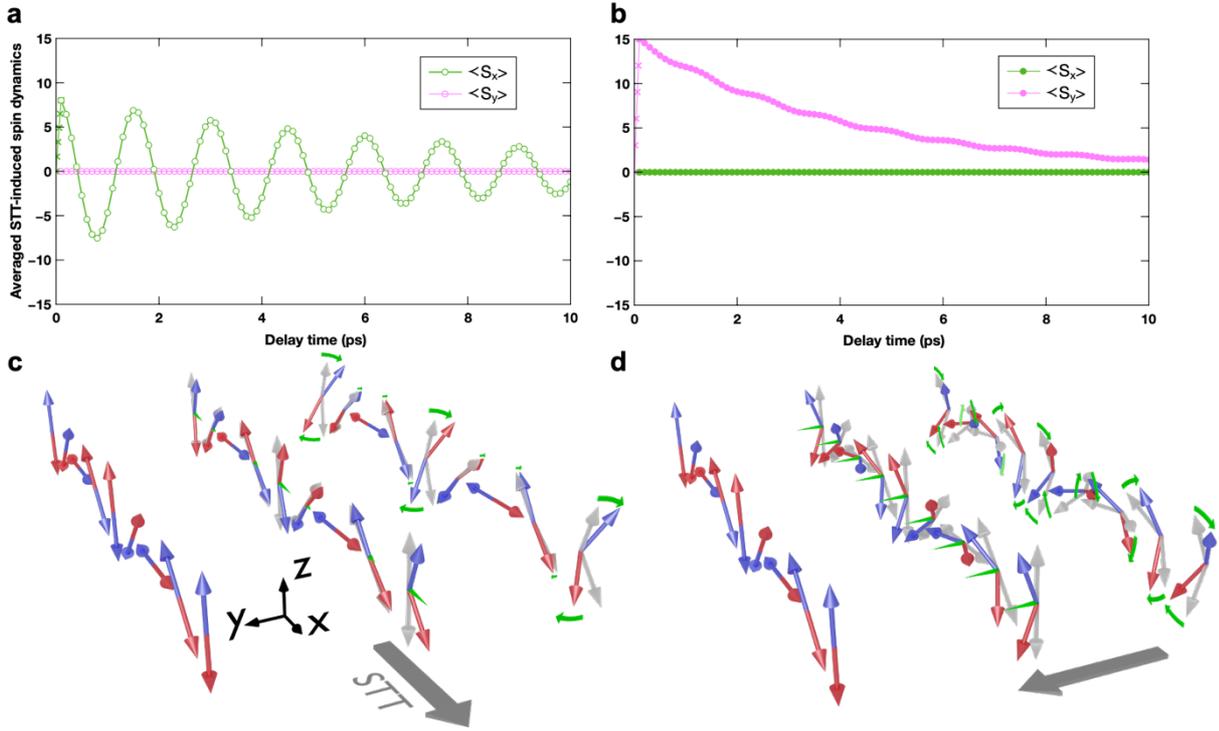

**Fig. 4 :** Atomic spin simulations of the STT-induced AF cycloid dynamics. a-b, Averaged magnetic moment components when the STT is applied along (a) and perpendicular (b) to the AF cycloidal plane. c-d, The scketches of their corresponding early stage spin dynamics for the STT along or perpendicular to the cycloidal plane, respectively.

Nonetheless, important additional insights can be obtained by performing atomic spin simulations using a Hamiltonian in which all relevant energy terms have been included (details can be found in Ref. [43]). A 64 nm pitch cycloid is obtained as the ground state in the proper $BiFeO_3$ anisotropy landscape. The AF cycloid dynamics is simulated after application of a 100 fs spin torque pulse mimicking the effect of light-induced spin accumulation at the CoFeB/$BiFeO_3$ interface during the ultra-fast ferromagnetic demagnetization process. The time evolution of the AF spin texture is assessed during the first 10 ps for a the spin torque applied along or perpendicular to the AF cycloidal plane. Figures 4a and 4b show the time-evolution of each component of magnetic moment averaged over one AF cycloid period. In both cases, a net magnetic moment is induced in the direction of the applied STT pulse, which is the signature of the injected spin-angular momentum. Noteworthy, the induced moment is halved for an STT applied in the AF cycloidal plane compared to when the STT is perpendicular to it. This can be naively expected as in this latter case, the STT is equivalent on all spins whereas in the cycloidal plane it has a sinusoidal dependence. In both cases, this initial period during the STT pulse, is followed by a clear oscillatory behaviour at the AF cycloid resonance frequency of 0.68 THz, though with a much larger amplitude when the STT is in the cycloidal plane. Note that the non-zero average magnetic moment induced in the direction of the applied STT shows a highly elliptical oscillation. The resulting overall STT-induced dynamics in the two geometries are fairly different (Fig. 4c-d). When the STT is along the AF cycloid (Fig. 4c), the spin dynamics is mostly located in the region where the spins are out of plane. However, with STT perpendicular to the AF cycloidal plane (Fig. 4d), all spins rotate around the induced magnetic moment, leading to an overall shift of the AF cycloid phase. While a quantitative comparison with the data is beyond the scope of this article, one can conclude that the simulations lead to a response frequency in $BiFeO_3$ fully consistent with that measured and a significant anisotropy in the amplitude, again consistent with the experiemental observation. This provides further evidence that the measured coherent dynamics in this insulating antiferromagnet is indeed triggered by spin transfer torque.

**Acknowledgements**

We are thankful for support from the French Agence Nationale de la Recherche (ANR) through the SPINUP (ANR-21-CE24-0026), THz-MUFINS (ANR-21-CE42-0030), TATOO (ANR-21-CE09-0033) projects and ESR/EquipEx+ program e-DIAMANT (Grant No. ANR-21-ESRE-0031), the European Union's Horizon 2020 research and innovation programme under the Grant Agreements No. 964931 (TSAR) and No. 866267 (EXAFONIS). This work is supported by a public grant overseen by the ANR as part of the "Investissements d'Avenir" program (Labex NanoSaclay, reference: ANR-10-LABX-0035). The Sesame Ile de France IMAGeSPIN project (No. EX039175) is also acknowledged.

1. NeéL, L. Antiferromagnetism and Ferrimagnetism. *Proc. Phys. Soc. A* **65**, 869 (1952).
2. Baltz, V. *et al.* Antiferromagnetic spintronics. *Rev. Mod. Phys.* **90**, 015005 (2018).
3. Jungfleisch, M. B., Zhang, W. & Hoffmann, A. Perspectives of antiferromagnetic spintronics. *Phys. Lett. A* **382**, 865 (2018).
4. Kittel, C. Theory of Antiferromagnetic Resonance. *Phys. Rev.* **82**, 565 (1951).
5. Keffer, F. & Kittel, C. Theory of Antiferromagnetic Resonance. *Phys. Rev.* **85**, 329 (1952).
6. Sievers, A. J. & Tinkham, M. Far Infrared Antiferromagnetic Resonance in MnO and NiO. *Phys. Rev.* **129**, 1566 (1963).
7. Kampfrath, T. *et al.* Coherent terahertz control of antiferromagnetic spin waves. *Nat. Photonics* **5**, 31 (2011).
8. Wang, H., Du, C., Hammel, P. C. & Yang, F. Antiferromagnonic Spin Transport from $Y_3Fe_5O_{12}$ into NiO. *Phys. Rev. Lett.* **113**, 097202 (2014).
9. Hahn, C. *et al.* Conduction of spin currents through insulating antiferromagnetic oxides. *EPL Europhys. Lett.* **108**, 57005 (2014).
10. Lin, W., Chen, K., Zhang, S. & Chien, C. L. Enhancement of Thermally Injected Spin Current through an Antiferromagnetic Insulator. *Phys. Rev. Lett.* **116**, 186601 (2016).
11. Lebrun, R. *et al.* Tunable long-distance spin transport in a crystalline antiferromagnetic iron oxide. *Nature* **561**, 222 (2018)
12. Ross, A. *et al.* Propagation Length of Antiferromagnetic Magnons Governed by Domain Configurations. *Nano Lett* **20**, 306 (2020).
13. Khymyn, R., Lisenkov, I., Tiberkevich, V. S., Slavin, A. N. & Ivanov, B. A. Transformation of spin current by antiferromagnetic insulators. *Phys. Rev. B* **93**, 224421 (2016).
14. Han, J. *et al.* Birefringence-like spin transport via linearly polarized antiferromagnetic magnons. *Nat. Nanotechnol.* **15**, 563 (2020).
15. Brataas, A., Tserkovnyak, Y., Bauer, G. E. W. & Kelly, P. J. Spin Pumping and Spin Transfer. In S. Maekawa, S. O. Valenzuela, E. Saitoh, & T. Kimura (Eds.), *Spin Current* (pp. 87-135). Oxford University Press.
16. Cheng, R., Xiao, J., Niu, Q. & Brataas, A. Spin Pumping and Spin-Transfer Torques in Antiferromagnets. *Phys. Rev. Lett.* **113**, 057601 (2014).
17. Li, J. *et al.* Spin current from sub-terahertz-generated antiferromagnetic magnons. *Nature* **578**, 70 (2020).
18. Vaidya, P. *et al.* Subterahertz spin pumping from an insulating antiferromagnet. *Science* **368**, 160 (2020).
19. Rongione, E. *et al.* Emission of coherent THz magnons in an antiferromagnetic insulator triggered by ultrafast spin–phonon interactions. *Nat. Commun.* **14**, 1818 (2023).
20. Beaurepaire, E., Merle, J.-C., Daunois, A. & Bigot, J.-Y. Ultrafast Spin Dynamics in Ferromagnetic Nickel. *Phys. Rev. Lett.* **76**, 4250 (1996).
21. Bigot, J.-Y & Vomir, M. Ultrafast magnetization dynamics of nanostructures. *Ann. Phys. (Berlin)* **525**, 2 (2013).
22. Kirilyuk, A., Kimel, A. V. & Rasing, T. Ultrafast optical manipulation of magnetic order. *Rev. Mod. Phys.* **82**, 2731 (2010).
23. Malinowski, G., Bergeard, N., Hehn, M. & Mangin, S. Hot-electron transport and ultrafast magnetization dynamics in magnetic multilayers and nanostructures following femtosecond laser pulse excitation. *Eur. Phys. J. B* **91**, 98 (2018).
24. Walowski, J. & Münzenberg, M. Perspective: Ultrafast magnetism and THz spintronics. *J. Appl. Phys.* **120**, 140901 (2016).
25. Koopmans, B. *et al.* Explaining the paradoxical diversity of ultrafast laser-induced demagnetization. *Nat. Mater.* **9**, 7 (2010).
26. Malinowski, G. *et al.* Control of speed and efficiency of ultrafast demagnetization by direct transfer of spin angular momentum. *Nat. Phys.* **4**, 855 (2008).
27. Battiato, M., Carva, K. & Oppeneer, P. M. Superdiffusive Spin Transport as a Mechanism of Ultrafast Demagnetization. *Phys. Rev. Lett.* **105**, 027203 (2010).
28. Melnikov, A. *et al.* Ultrafast Transport of Laser-Excited Spin-Polarized Carriers in Au/Fe/MgO(001). *Phys. Rev. Lett.* **107**, 076601 (2011).
29. Alekhin, A. *et al.* Femtosecond Spin Current Pulses Generated by the Nonthermal Spin-Dependent Seebeck Effect and Interacting with Ferromagnets in Spin Valves. *Phys. Rev. Lett.* **119**, 017202 (2017).
30. Razdolski, I. *et al.* Nanoscale interface confinement of ultrafast spin transfer torque driving non-uniform spin dynamics. *Nat. Commun.* **8**, 15007 (2017).
31. Kampfrath, T. *et al.* Terahertz spin current pulses controlled by magnetic heterostructures. *Nat. Nanotechnol.* **8**, 256 (2013).
32. Seifert, T. *et al.* Efficient metallic spintronic emitters of ultrabroadband terahertz radiation. *Nat. Photonics* **10**, 483 (2016).
33. Papaioannou, E. Th. & Beigang, R. THz spintronic emitters: a review on achievements and future challenges. *Nanophotonics* **10**, 1243 (2021).
34. Burns, S. R., Paull, O., Juraszek, J., Nagarajan, V. & Sando, D. The Experimentalist's Guide to the Cycloid, or Noncollinear Antiferromagnetism in Epitaxial $BiFeO_3$. *Adv. Mater.* **32**, 2003711 (2020).
35. Cazayous, M. *et al.* Possible Observation of Cycloidal Electromagnons in $BiFeO_3$. *Phys. Rev. Lett.* **101**, 037601 (2008).
36. Sando, D. *et al.* Crafting the magnonic and spintronic response of $BiFeO_3$ films by epitaxial strain. *Nat. Mater.* 9 (2013).
37. Haykal, A. *et al.* Antiferromagnetic textures in BiFeO3 controlled by strain and electric field. *Nat. Commun.* **11**, 1704 (2020).
38. Dufour, P. *et al.* Onset of Multiferroicity in Prototypical Single-Spin Cycloid $BiFeO_3$ Thin Films. *Nano Lett.* **23**, 9073 (2023).
39. Dornes, C. *et al.* The ultrafast Einstein–de Haas effect. *Nature* **565**, 209 (2019).
40. Frangou, L. *et al.* Enhanced Spin Pumping Efficiency in Antiferromagnetic IrMn Thin Films around the Magnetic Phase Transition. *Phys. Rev. Lett.* **116**, 077203 (2016).
41. Saenrang, W. Deterministic and robust room-temperature exchange coupling in monodomain multiferroic $BiFeO_3$ heterostructures. *Nat. Commun.* **8**, 1583 (2017).
.42. Elzo, M. *et al.* Coupling between an incommensurate antiferromagnetic structure and a soft ferromagnet in the archetype multiferroic $BiFeO_3$/cobalt system. *Phys. Rev. B* **91**, 014402 (2015).
43. Li, Z. *et al.* Multiferroic skyrmions in $BiFeO_3$. *Phys. Rev. Res.* **5**, 043109 (2023).

**METHODS:**

**Sample preparation:**

A 66 nm thick $BiFeO_3$ layer was grown by pulsed laser deposition on top of a $SrRuO_3$-buffered (011)-oriented $DyScO_3$ orthorhombic substrate ([111] pseudo-cubic axis)[38]. Such a substrate promotes the growth of $BiFeO_3$ along its polar axis, resulting in a single ferroelectric domain with a purely out-of-plane polarization pointing downward, as checked with a combination of advanced X-ray diffraction and piezoresponse force microscopy experiments[38]. Moreover, the anisotropic in-plane epitaxial strain lifts the degeneracy between the AF cycloidal variants. A single AF domain results with its cycloidal propagation direction, parallel to the a axis of the $DyScO_3$ substrate. The antiferromagnetic spin textures of $BiFeO_3$ were imaged (Fig. 1b) using a commercial scanning NV magnetometer (ProteusQ™, Qnami AG) operated under ambient conditions. The scanning tip is a commercial all-diamond probe with a single NV defect at its apex, integrated on a quartz tuning fork (Quantilever™ MX, Qnami AG). The diamond tip is integrated into a tuning-fork based atomic force microscope combined with a confocal microscope optimized for single NV defect spectroscopy. Subsequently a 5 nm thick $Co_{40}Fe_{40}B_{20}$ layer, capped by 2 nm of Al, was simultaneously grown by sputtering on top of the $BiFeO_3$ film and on a bare (011) $DyScO_3$ substrate for reference.

**Time-resolved magneto-optical experiments:**

The laser pulses are generated by a 1 kHz Ti:sapphire amplifier with a pulse duration of 65 fs. The pump (800 nm wavelength) is focused on a 300 μm spot with a 60° incident angle. The probe (400 nm wavelength) is focused on a 22 μm spot at a 44° incidence angle with respect to the sample plane. After reflection on the sample, the ultrafast change in light characteristics is measured by a standard polarization bridge. The probe fluence is about 75 µJ/cm² while the pump fluence was varied between typically 1.5 and 9 mJ/cm².

In order to extract the relevant parameters, the experimental data are fitted with the following phenomenological model:

$$m(t) = \Theta(t) \otimes H(t) \times \left(A_D e^{-\frac{t}{\tau_D}} + A_R e^{-\frac{t}{\tau_R}} + \cdots\right)$$

Where $m(t)$ is the magnetization, $H(t)$ is the Heaviside step function, and several exponential terms are assessing the characteristics of the various demagnetization and remagnetization processes. The whole magnetization dynamics profile is convoluted by 2 gaussian curves ($\Theta(t)$) in order to reproduce the effect of the experimental time resolution due to both the pump and probe temporal profiles.

**Atomic spin dynamics simulations:**

On a general viewpoint, simulating AF properties requires to consider every single spins. Atomic spin dynamics simulations have therefore been carried out using a GPU-based homemade code. The ground state of our [111] oriented BiFeO3 is found to be a 64 nm pitched single AF cycloid along the [1-10] direction. Here, we consider a 64 nm x 1.94 nm x 1.38 nm simulated area, in a frame defined by the [1-10], [11-2] ,[111] directions, with periodic boundary conditions in all directions. This reproduces a BiFeO$_3$ layer hosting a single cycloid with the ferroelectric polarization along the [111] direction.To simulate the effect of an ultrafast spin current, damping-like STT is applied for the first 100 fs on the whole simulated area. Note that as in the experiments the BiFeO$_3$ layer is 66 nm thick, the STT is expected to be applied only on the few first nanometers from the interface. Nontheless, the simulations should capture the main features of the STT-induced dynamics of the AF cycloid. More modelling details can be found in Z. Li et al[43].